\documentclass[smallextended]{svjour3}
\pdfoutput=1

\usepackage{graphicx,amsmath,amssymb,amsbsy,bm,color,natbib,pgfplots,mathrsfs,hyperref}
\usepackage{mathptmx}
\usepackage[T1]{fontenc}

\hypersetup{colorlinks,%
                 linkcolor=blue,%
                 citecolor=blue,%
                 pdftex}

\let\le=\leqslant  
\let\ge=\geqslant

\providecommand\etal{\textit{et al.}}
\providecommand\mathsfi[1]{\mathbf{#1}}

\providecommand\eps\varepsilon
\providecommand\del\delta
\providecommand\p\partial

\renewcommand\i{\text{i}}

\begin{document}

\title{Artificial eigenmodes in truncated flow domains}
\author{Lutz Lesshafft
%\thanks{\href{mailto:lesshafft@ladhyx.polytechnique.fr}{lesshafft@ladhyx.polytechnique.fr}} 
}
\institute{Laboratoire d'Hydrodynamique, CNRS/\'Ecole polytechnique,
91128 Palaiseau, France\newline \email{lesshafft@ladhyx.polytechnique.fr}}

\date{\today}
\maketitle

\begin{abstract}
Whenever linear eigenmodes of open flows are computed on a numerical domain that is truncated in the streamwise direction, artificial boundary conditions may give rise to spurious pressure signals that are capable of providing unwanted perturbation feedback to upstream locations. The manifestation of such feedback in the eigenmode spectrum is analysed here for two simple configurations. First, explicitly prescribed feedback in a Ginzburg--Landau model is shown to produce a spurious eigenmode branch, named the `arc branch', that strongly resembles a characteristic family of eigenmodes typically present in open shear flow calculations. Second, corresponding mode branches in the global spectrum of an incompressible parallel jet in a truncated domain are examined. It is demonstrated that these eigenmodes of the numerical model depend on the presence of spurious forcing of a local $k^+$ instability wave at the inflow, caused by pressure signals that appear to be generated at the outflow. Multiple local $k^+$ branches result in multiple global eigenmode branches. For the particular boundary treatment chosen here, the strength of the pressure feedback from the outflow towards the inflow boundary is found to decay with the cube of the numerical domain length. It is concluded that arc-branch eigenmodes are artifacts of domain truncation, with limited value for physical analysis. It is demonstrated, for the example of a non-parallel jet, how spurious feedback may be reduced by an absorbing layer near the outflow boundary.
\end{abstract}

%\begin{keywords}global modes, domain truncation, nonreflecting\end{keywords}

\section{Introduction}

Linear instability analysis of open flows today is commonly carried out in a so-called `global' framework, where at least two non-homogeneous spatial directions of a steady base state are numerically resolved. In contrast to `local' analysis, where the base state is assumed to be parallel, and unbounded in the flow direction, a global discretisation of an open flow problem in a truncated numerical domain necessitates the formulation of artificial streamwise boundary conditions for flow perturbations. The question then arises in how far such boundary conditions influence the instability dynamics of the truncated flow system. 

\begin{figure}
\label{fig:spectrumjet}
\centering
\includegraphics[width=0.8\textwidth]{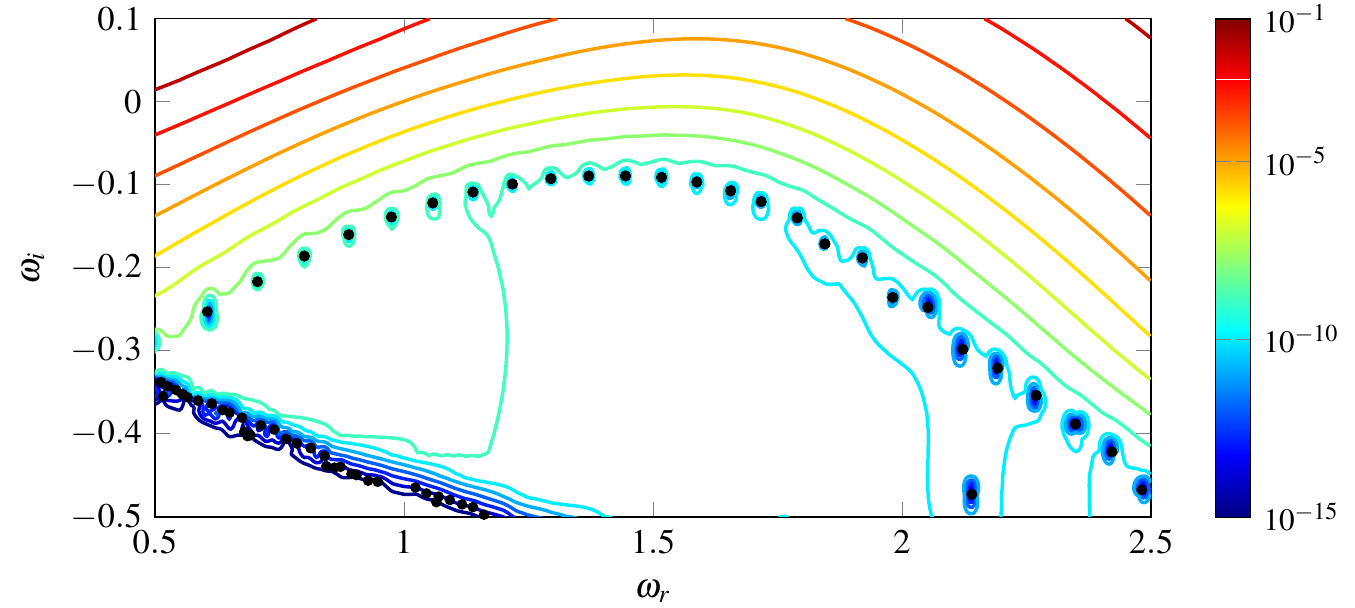}
\caption{Spectrum and pseudospectrum of a jet at $Re=1000$, showing a typical arc branch. Eigenmodes are drawn as black symbols. The definition of $\epsilon$ values, shown as colour contours, is given in Sect.~\ref{sec:GL}. The figure is made from the data of Coenen \etal~\cite{Coenen2017}.}
\end{figure}

This paper investigates the effect of spurious pressure feedback, due to domain truncation, on the eigenmode spectrum of incompressible open flow problems. The investigation is motivated by observations made in recent linear instability studies of jet flows \cite{Garnaud:2013p1182,Coenen2017}, where a prominent family of eigenmodes (black symbols in Fig.~\ref{fig:spectrumjet}), referred to as the `arc branch' from here on, 
was found to present features that suggest a resonance between the inflow and outflow boundaries. Such branches are in fact ubiquitous in many, if not all, global spectra of truncated open shear flows found in the literature: boundary layers \citep{Ehrenstein:2005p1026,Ehrenstein:2008p1126,Akervik:2008p1176}, cylinder wakes \citep{Sipp2007,Marquet:2008p1127},  jets \citep{Nichols:2011p1075,Garnaud:2013p1182}, plumes \citep{ChakriPhD}, three-dimensional boundary layers with roughness elements \citep{Loiseau2014investigation, Kurz2016} --- all these and many others present similar characteristic branches of eigenvalues that are often described as being highly dependent on the type or position of outflow boundary conditions. Typically, no convergence with respect to the length of the numerical box can be attained for such modes.

Ehrenstein \& Gallaire \cite{Ehrenstein:2005p1026} remark on the resemblance between arc-branch-type global eigenmode structures obtained for a flat-plate boundary layer and spatial modes as found in a local analysis. \AA{}kervik \etal~\cite{Akervik:2008p1176} document the dependence of boundary layer eigenvalues on the type of outflow boundary conditions. Heaton \etal~\cite{heaton} characterise arc branch modes in the spectrum of a Batchelor vortex as artifacts, speculating that these arise from the limited precision of their numerical scheme, not from domain truncation. Cerqueira \& Sipp \cite{cerqueira} demonstrate that such precision errors indeed lead to spurious `quasi-eigenmodes', but that these differ from the arc branch. In their analysis, quasi-eigenmodes appear in regions of the complex frequency plane where pseudospectrum $\epsilon$ values \cite{TrefethenBook} are very small, below approximately $10^{-12}$. These modes, also visible in the lower left corner of Fig.~\ref{fig:spectrumjet}, are very sensitive to the numerical scheme, to mesh refinement and to the eigenvalue shift parameter. Arc branch modes are found to be robust with respect to those details\cite{cerqueira}, but strongly dependent on the numerical domain length.
Coenen \etal \cite{Coenen2017} show that arc branch eigenfunctions of a jet are characterised by an integer number of wavelengths between the inflow and outflow boundaries, and they suggest an analogy with acoustic modes in a pipe of finite length. This analogy implies that arc branch modes are the result of unwanted resonance between the numerical boundaries, potentially leading to a spurious  instability of the numerical system. The present study expands on all these observations, and it aims at a detailed characterisation of unphysical resonance due to imperfect boundary conditions in open shear flow calculations.

The possibility that global instability in truncated systems may be brought about by spurious pressure feedback from boundary conditions was probably first described by Buell \& Huerre\cite{BuellHuerre}. In direct numerical simulations of perturbations in a mixing layer, unstable perturbation growth was observed at long times, in a configuration that was only convectively unstable in a local sense. Such behaviour is inconsistent with the interpretation of local convective instability, and it was demonstrated to be caused by unphysical pressure perturbations emanating from the outflow boundary, which in turn provoked the formation of vortical perturbations at the inflow boundary. Chomaz \cite[][section 3.2.2]{Chomaz2005} interprets pressure feedback as a non-local operator variation, arguing that strong non-normality in the spectrum of convection-dominated flows is likely to induce a high sensitivity of global eigenmodes with respect to such feedback. 

While the spurious generation of acoustic pressure waves from artificial boundary conditions is an important and much-discussed problem in compressible flow simulations, especially those that aim to accurately capture the acoustic radiation from shear flows \citep{C04}, the question of how such artifacts may affect the global stability behaviour remains largely unexplored, both in compressible and in incompressible configurations. 

The problem is approached here in the following manner. Section 2 presents a global instability analysis of the Ginzburg--Landau equation with explicit feedback from a downstream sensor to an upstream actuator, in order to examine the effect of such feedback on the eigenvalue spectrum in a controlled setting. This model study fully describes the suspected mechanism behind arc branch modes. In Sect.~3 the analysis is extended to a parallel jet flow in a finite-size numerical domain. Global eigenmodes are projected onto their local counterparts, the global pressure field is examined, and spurious feedback effects are analysed. The question how spurious feedback may be reduced in a practical manner is addressed in Sect.~4, for the example of a spatially developing jet.

\section{A Ginzburg--Landau model problem}

\label{sec:GL}

The hypothesis put forward by Coenen \etal \cite{Coenen2017}, that the arc
branch may be the manifestation of a non-physical upstream scattering of
perturbations from the outflow boundary, is first investigated with the help of a
simplified model. The linear Ginzburg--Landau equation is written as
\begin{equation}
\label{eqn:GL}
\partial_t \psi =
- U\partial_x \psi + \mu(x) \psi + \gamma\partial_{xx} \psi + f(x,\psi) \, .
\end{equation}
The complex scalar variable $\psi(x,t)$ is a function of time $t$ and of one
single spatial coordinate $x$. Constant parameters $U=6$ and $\gamma=1-\i$ are
chosen \emph{ad hoc}, whereas the coefficient $\mu$ varies linearly in $x$ as
\begin{equation}
\mu(x) = \frac{U^2}{8}\left( 1- \frac{x}{20} \right).
\end{equation}
With this particular variation of $\mu$, the system is marginally absolutely
unstable at $x=0$, convectively unstable for $0<x<20$, and stable for $x>20$,
all in a local sense. Ginzburg--Landau systems of this form, with linearly
decreasing $\mu(x)$ and without feedback, $f\equiv 0$, have been extensively used to model the
global instability behaviour of spatially developing flows in semi-infinite
domains \citep[see for instance][]{Couairon97}. The instability characteristics mimic,
in a very simple and qualitative manner, those of a spreading jet. However, the system (\ref{eqn:GL}) is not intended here to predict or reproduce the dynamics of any specific flow. 

A forcing term $f(x,\psi)$ is added in (\ref{eqn:GL}) in order to provide
an explicit closed-loop forcing between the upstream and the downstream end of
the flow domain. Some aspects of closed-loop forcing in the Ginzburg--Landau equation are discussed by Chomaz \cite{Chomaz2005}; in the present context, it is used to model a suspected spurious feedback in global shear flow
computations. Taken to be of the form
\begin{equation}
f(x,\psi) = C \exp\left( -\frac{(x-x_a)^2}{0.1^2} \right) \psi(x_s),
\end{equation}
a forcing proportional to $\psi(x_s)$ is applied in a close vicinity of $x_a$,
such that a feedback loop is established between a (downstream) sensor location
$x_s$ and an (upstream) actuator location $x_a$. The complex coefficient $C$
governs the amplitude and phase of the feedback, and the Gaussian spreading
around $x_a$ is introduced for reasons of numerical resolution.

Equation (\ref{eqn:GL}) is discretised on an interval $0\le x \le 40$, with a
step size $\Delta x=0.1$, using an upwind-biased seven-point finite difference
stencil for the spatial derivatives. A homogeneous Dirichlet
boundary condition for $\psi$ is imposed at the upstream boundary, consistent
with typical jet conditions. Actuator and sensor locations are chosen close to the
boundaries, at $x_a=1$ and $x_s=39$, where spatial derivatives are well resolved.

\begin{figure}
\centering
\includegraphics[width=\textwidth]{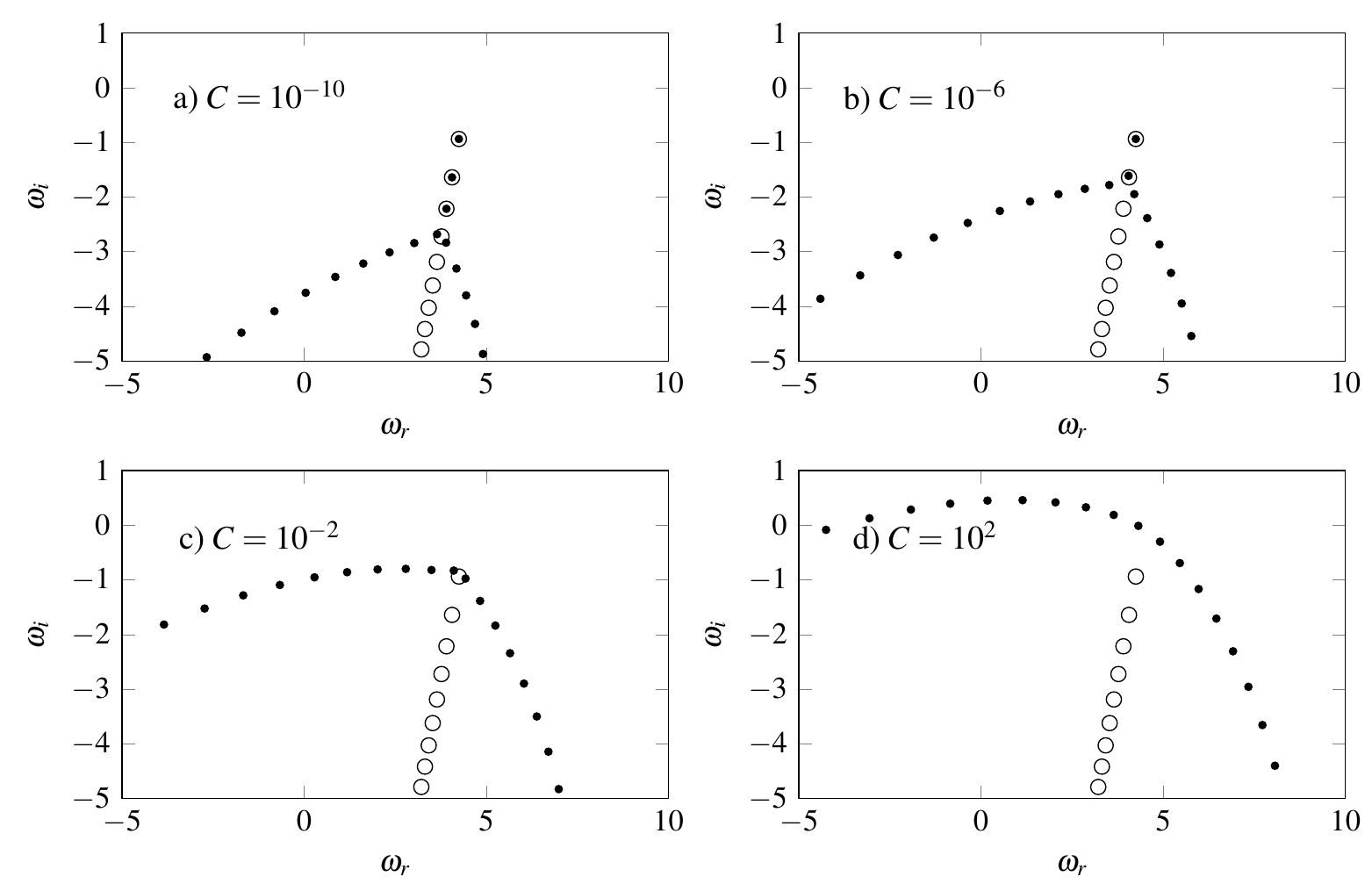}
\caption{%
Spectra of the Ginzburg--Landau
equation (\ref{eqn:GL}) with various feedback coefficients $C$.  Eigenvalues with 
closed-loop feedback ($\bullet$), and analytical eigenvalues (\ref{eqn:anEV}) of the 
case without feedback ({\large$\circ$}).}
\label{fig:GL_spectrum}
\end{figure}

Temporal eigenmodes of (\ref{eqn:GL}) are sought in the form $\psi(x,t)=\hat{\psi}(x)\, \text{e}^{-i\omega t}$. The eigenvalues of the system without feedback, $C=0$, are known analytically 
to be
\begin{equation}
\label{eqn:anEV}
\omega_n = i\left\{\frac{U^2}{8} - \frac{U^2}{4\gamma}
+ \gamma^\frac{1}{3}\frac{U^\frac{4}{3}}{160^\frac{2}{3}} \zeta_n \right\},
\end{equation}
where $\zeta_n$ is the $n^{th}$ root of the Airy function \cite{CHR88}.
These values are represented in Fig.~\ref{fig:GL_spectrum} as circles. Each
frame~\ref{fig:GL_spectrum}(\textit{a}--\textit{d}) also displays the
numerically computed eigenvalues of systems with feedback, shown as 
bullet symbols, for different non-zero values of the coefficient $C$. Already with
very low-level feedback, $C=10^{-10}$, the spectrum is clearly affected: only
the leading three eigenvalues of the unforced system are recovered, and the
lower part of the spectrum is replaced with two new stable branches. These
branches, named \emph{feedback branches} in the following, move upward in the
complex $\omega$ plane as the feedback coefficient is increased, masking more
and more of the original unforced eigenvalues. Note that those affected
original eigenvalues are not merely altered by the feedback, but they rather
disappear abruptly from the spectrum as they fall below the new
branches. 

The spectra obtained with $C=10^{-6}$ and $10^{-2}$ resemble those of pure
helium jets, as shown in 
%figure 1 of \cite{Lesshafft2014},
figure 6 of \cite{Coenen2017}, 
and those of cylinder wakes obtained by Marquet \etal \cite[][their figure 16]{Marquet:2008p1127}. The least stable original eigenvalue lies just
above the feedback branches, apparently unaffected. In the strongest feedback
case, $C=10^2$, the feedback branches have merged into one, overarching the
entire spectrum of the feedback-free system, which has altogether disappeared.
This picture (Fig.~\ref{fig:GL_spectrum}$d$) resembles the spectrum of
the slowly developing jet of Coenen \etal \cite{Coenen2017}, 
%the laminar constant-density jet of \cite{Garnaud:2013p1182},
reproduced in our Fig.~\ref{fig:spectrumjet}.

\begin{figure}
\centering
\includegraphics[width=\textwidth]{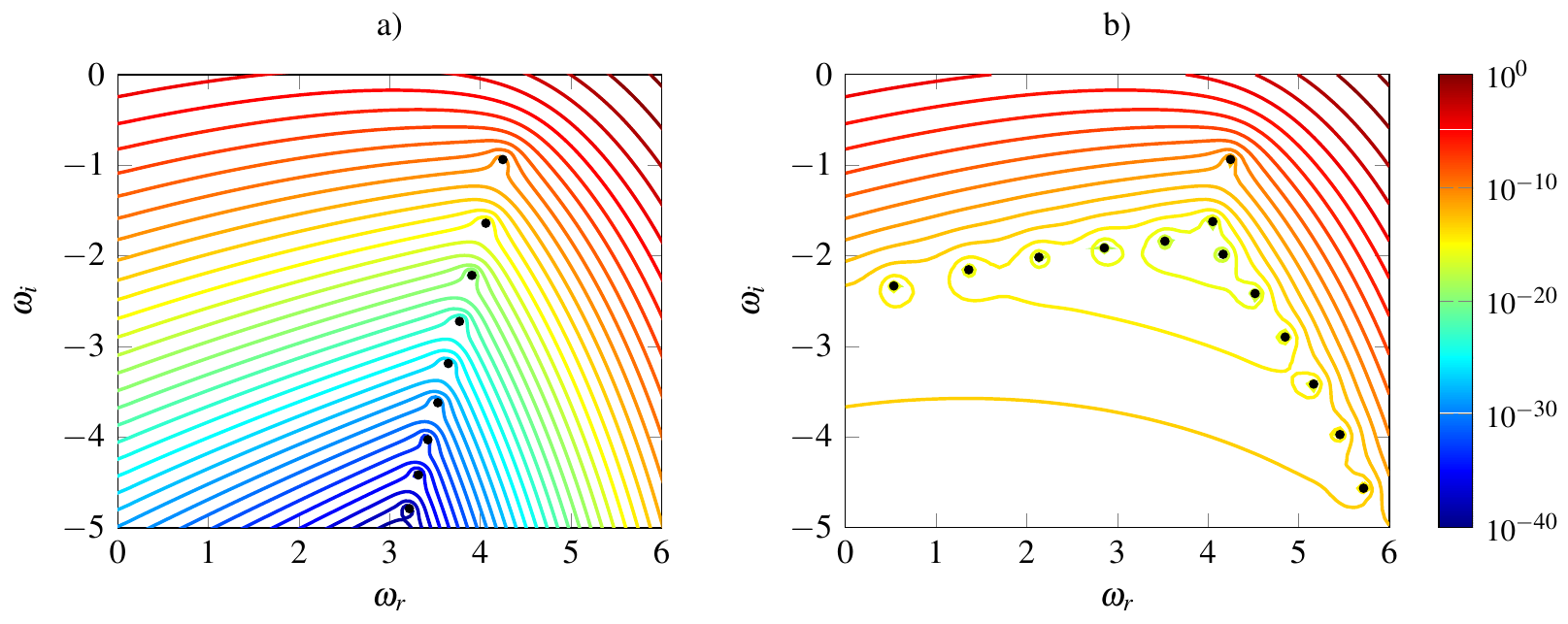}
\caption{%
Pseudospectra of the Ginzburg--Landau equation; ($a$) without feedback, $C=0$, and 
($b$) with feedback, $C=10^{-6}$. Contours of $\epsilon$, defined in (\ref{eqn:pseudospec}), are represented 
in logarithmic colour scale.}
\label{fig:GLpseudo}
\end{figure}

The similarity between the feedback branches in the present model and the arc branch in the jet spectrum (Fig.~\ref{fig:spectrumjet}) is also manifest in the pseudospectra. 
If feedback is thought of as a variation of the operator \cite{Chomaz2005}, it should be possible to relate feedback-induced eigenvalues to the pseudospectrum.  
The pseudospectrum is defined here by the spectral norm of the resolvent operator, 
\begin{equation}
\label{eqn:pseudospec}
\partial_t \psi = \mathsfi{L}\psi \quad \rightarrow \quad \|(\i\omega\mathsfi{I}+\mathsfi{L})^{-1}\| = \epsilon^{-1},
\end{equation}
at any complex frequency $\omega$ \cite{TrefethenBook}.  Pseudospectrum
$\epsilon$-contours of the Ginzburg--Landau equation are compared between the unforced
setting with $C=0$, Fig.~\ref{fig:GLpseudo}($a$), and the forced setting
with $C=10^{-6}$, Fig.~\ref{fig:GLpseudo}($b$). It is seen that the feedback
branches align closely with an isocontour of the feedback-free pseudospectrum.
This criterion also applies to all other cases displayed in Fig.~\ref{fig:GL_spectrum}, with different values of $C$, and it is fully consistent with observations made in three flow configurations \cite{heaton,cerqueira,Coenen2017}. Furthermore, the pseudospectrum of the system with feedback is identical to that of the system without feedback \emph{above}
the feedback branches, whereas the pseudospectrum is flattened, nearly
constant, \emph{below} the feedback branches. The same behaviour is found in the jet pseudospectrum shown in Fig.~\ref{fig:spectrumjet}.

\begin{figure}
\centering
\includegraphics[width=0.9\textwidth]{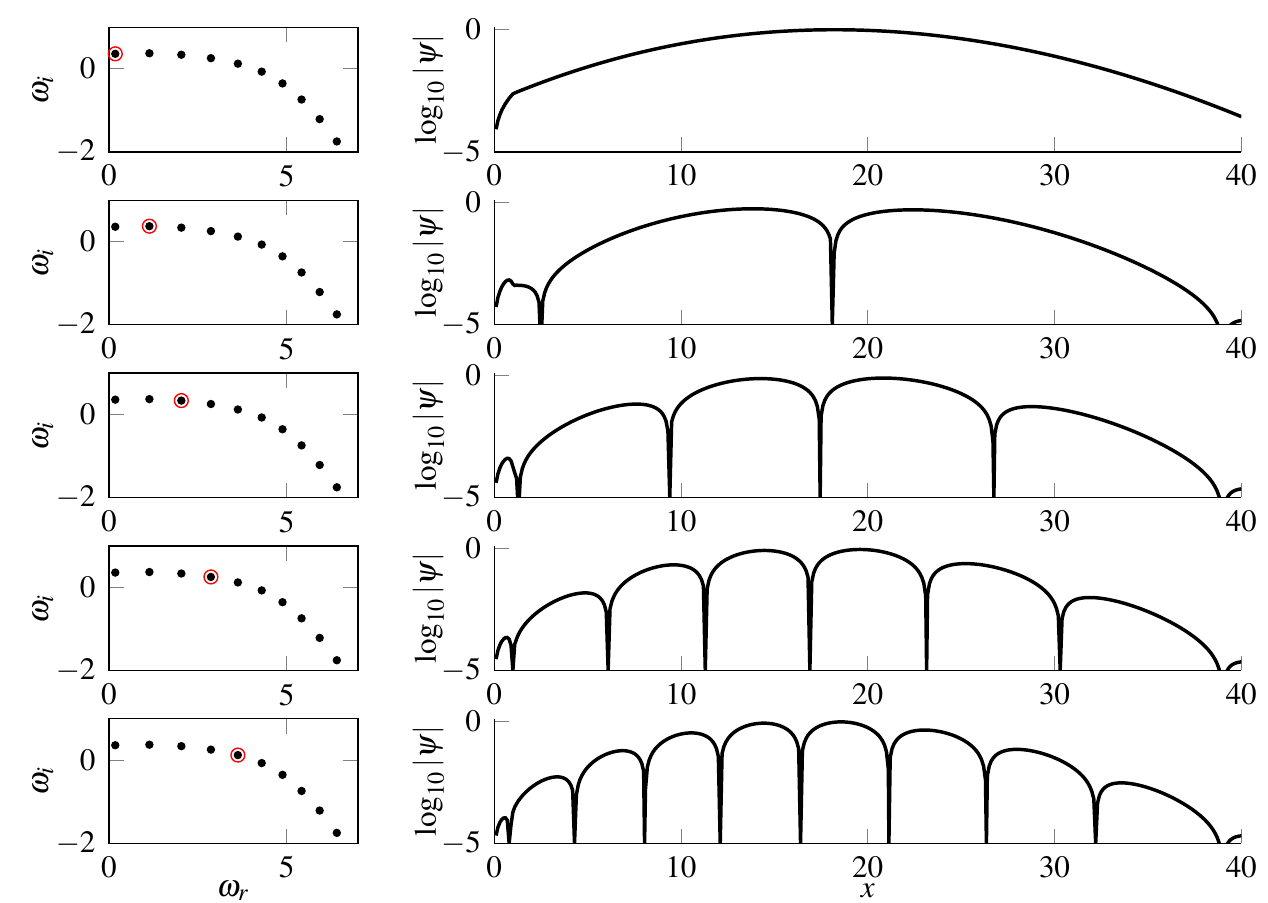}
\caption{%
Eigenfunctions of the feedback branch for $C=10^2$. Left column: position of
the eigenvalue in the spectrum; right column: corresponding eigenfunction
$\log_{10}|Re(\psi)|$ along $x$. Zero wavelengths are present in the top frame,
one in the second, and so forth up to four wavelengths in the bottom frame. The series continues further along the branch.}
\label{fig:GLwave}
\end{figure}

Finally, the discrete distribution of feedback modes along the branch is
investigated. The strong feedback case $C=10^2$ is considered for illustration.
Eigenfunctions of successive feedback modes are presented in figure
\ref{fig:GLwave}, 
analogous to the representation of jet results in figure 5 of \cite{Coenen2017}. The absolute value of $\psi$ is traced in logarithmic scale as a function of $x$. 
%where the absolute value of $\psi$ is traced in logarithmic scale as a function of $x$. 
%
The phase of $\psi$ is always chosen such that the real part of $\psi$ is zero in the sensor location $x_s=39$, with the exception of the first mode
(Fig.~\ref{fig:GLwave}$a$), which has no interior wave nodes. Only positive real
frequencies are considered.

Clearly, the eigenfunctions are characterised by integer numbers of wavelengths
between the actuator and sensor locations, starting from zero wavelengths in
the case of the lowest real frequency (Fig.~\ref{fig:GLwave}$a$, only
amplitude variations but approximately constant phase), to one wavelength in
Fig.~\ref{fig:GLwave}$(b)$, and so forth with a continuously increasing
count. The series could be continued further along the entire arc branch. 
The present model seems to reproduce very well the characteristics of arc branch eigenfunctions, as displayed in figure 5 of \cite{Coenen2017}.

\begin{figure}
\centering
\includegraphics[width=0.6\textwidth]{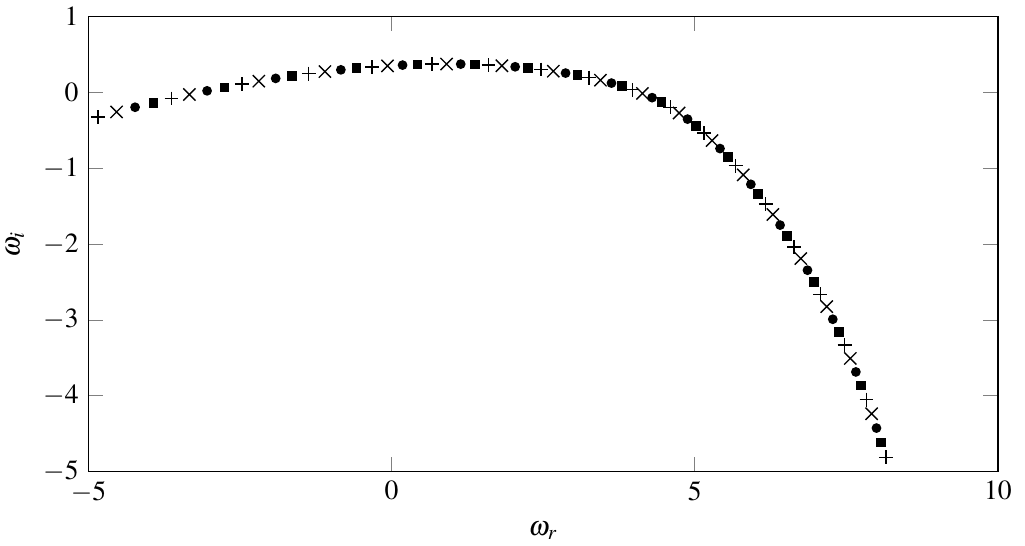}
\caption{%
Eigenvalues of the feedback branch for $|C|=10^2$ with varying phase:\newline
($\bullet$) $C=10^2$, ($\times$) $C=10^2 \i$, (+) $C=-10^2$,
(\protect\raisebox{1pt}{\tiny$\blacksquare$})  $C=-10^2 \i$.}
\label{fig:GLphase}
\end{figure}

The position and spacing of eigenvalues along the branch curve, fixed
approximately by an isocontour of the pseudospectrum, appears to be determined
by a fitting phase relation between $\psi$ at the actuator location and the
applied feedback. To further illustrate this mechanism, the phase of the
feedback coefficient $C$ is varied. Figure \ref{fig:GLphase} shows the arc
branch as it is obtained with values $C=10^2$ (bullets, same as before),
$C=10^2 \i$ (crosses), $C=-10^2$ (plus signs) and $C=-10^2 \i$ (squares). All
symbols fall onto the same curve, confirming that the position of feedback
eigenmodes, which represent singularities in the pseudospectrum, is fixed by
the phase of the feedback relation, such that resonance can occur between both
ends of the loop.

%%%%%%%%%%%%%%%%%%%%%%%%%%%%%%%%%%%%%%%%%%%%%
%%%%%%%%%%%%%%%%%%%%%%%%%%%%%%%%%%%%%%%%%%%%%

\section{A parallel jet}
\label{sec:jet}

The effect that imperfect numerical boundary conditions can have on eigenmode computations in a two-dimensional flow domain is now investigated for the case of axisymmetric perturbations in a parallel round jet, governed by the incompressible axisymmetric Navier--Stokes equations,
\begin{subequations}
\label{eqn:nonlineqns}
\begin{align}
0 &= \p_r u_r + \frac{u_r}{r} + \p_x u_x, \\
\p_t u_r &= - u_r \p_r u_r - u_x \p_x u_r - \p_r p + \frac{1}{Re}\left( \p_{rr}u_r + \frac{\p_r u_r}{r} -\frac{u_r}{r^2} + \p_{xx}u_r \right) ,\\
\p_t u_x &= - u_r \p_r u_x - u_x \p_x u_x- \p_x p + \frac{1}{Re}\left( \p_{rr}u_x + \frac{\p_r u_x}{r}  + \p_{xx}u_x \right) .
\end{align}
\end{subequations}
Boundary conditions are prescribed as
\begin{subequations}
	\label{eqn:bcs}
\begin{align}
&u_r=u_x=0 \quad \text{at }x=0, \\
&Re^{-1}\p_x u_x - p = 0 \quad \text{at }x=x_{max}, \\
&u_r = \p_r u_x  = 0 \quad \text{at }r=0\text{~and~}r=r_{max}.
\end{align}
\end{subequations}
The stress-free outflow condition is a common and convenient choice for finite-element computations. Jet radius and centreline velocity are the dimensional length scales in this formulation, and the Reynolds number is chosen as $Re=100$. 

The advantage of using a parallel base flow profile is that the global results can be rigorously compared with local instability properties. The standard analytical model of Michalke \cite{M84} is adopted,
\begin{equation}
\label{eqn:baseflow}
U(r) = \frac{1}{2}\left(1 + \tanh\left[\frac{1}{4\theta}\left( \frac{1}{r} - r\right)\right]\right),
\end{equation}
and a momentum thickness $\theta=0.1$ is chosen for this example. 

The axisymmetric perturbation equations that follow from linearization of \eqref{eqn:nonlineqns} about the base flow (\ref{eqn:baseflow}) are discretized with finite elements on a domain of length $x_{max}=20$ and radial extent $r_{max}=50$, using the FreeFEM++ software. This domain is resolved with 80 equidistant elements in $x$, and with 360 non-equidistant elements in $r$. The equations are solved for eigenmodes $[u_r,u_x,p]^T(r,x,t) = [\hat{u}_r, \hat{u}_x,\hat{p}]^T (r,x) \exp (-\i\omega t)$, where the eigenvalue $\omega = \omega_r +\i\omega_i$ contains the angular frequency $\omega_r$ and the temporal growth rate $\omega_i$.

\begin{figure}
\centering
\includegraphics[width=0.6\textwidth]{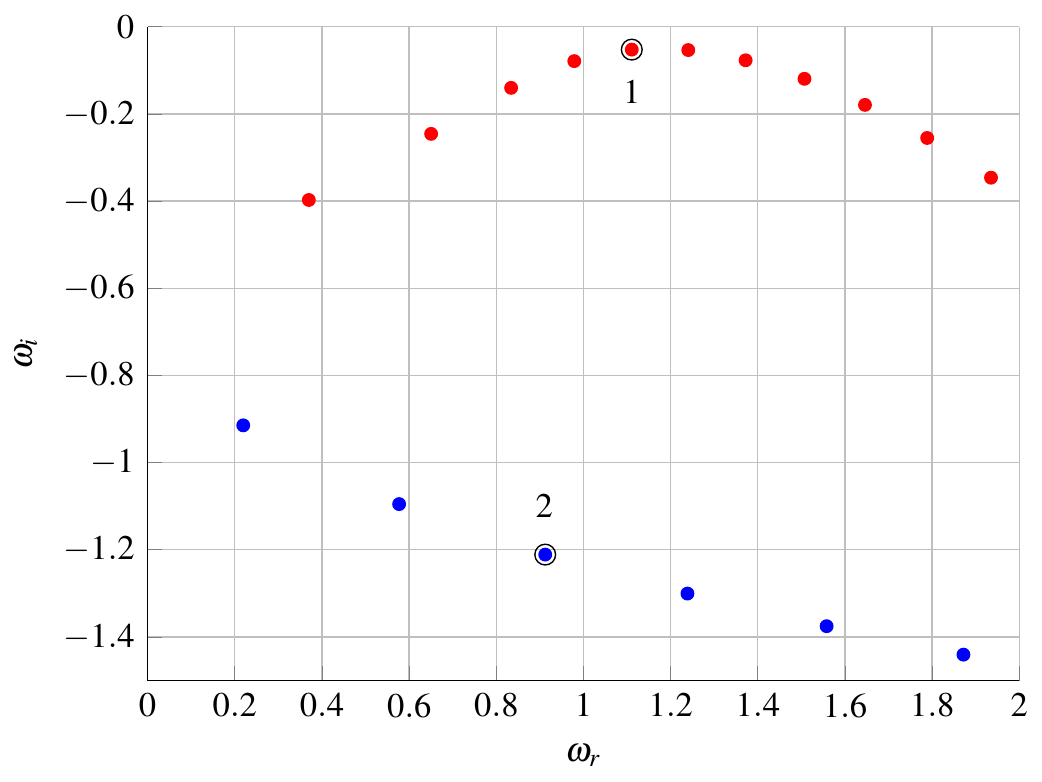}
\caption{Spectrum of the parallel jet in a finite domain with inflow and outflow boundary conditions. Red: arc branch; blue: lower branch. Labelled modes are discussed in the following.}
\label{fig:globalspectrum}
\end{figure}

The global spectrum is shown in Fig.~\ref{fig:globalspectrum}. It features a clean upper arc branch (red) with maximum growth rate at $\omega_1=1.111 - 0.052\i$. This mode, labelled `1', is chosen for further analysis. A lower branch (blue) is also present, from which the mode labelled `2' with $\omega_2=0.912 - 1.211\i$ will be examined. Similar lower branches are visible in wake spectra \cite{Sipp2007} and in boundary-layer calculations \cite{Akervik:2008p1176}.

\subsection{Projection onto local instability modes}
\label{sec:local}

For comparison, the corresponding local instability problem, for a domain of infinite extent $x\in (-\infty,\infty)$, is solved with a standard Chebyshev collocation technique on a staggered grid \cite{Khorrami:1991p1179}, using a coordinate transformation adapted for jet profiles \cite{Lesshafft:2007p54}. Identical radial collocation point distributions are used in the global and local computations, such as to eliminate the need for interpolation. Consistent boundary conditions (\ref{eqn:bcs}c) are imposed in the local problem.

Spatial local instability modes are computed for the $\omega$ values corresponding to the global modes labelled in Fig.~\ref{fig:globalspectrum}. The flow is convectively unstable, with an absolute frequency $\omega_0=1.074-0.286\i$. Both direct and adjoint modes are solved for. The adjoint local modes represent the dual basis associated with the set of direct modes, and they serve for projecting the spatial structure of the global mode onto the local direct modes. 

This projection is carried out in the following way, similar to the procedure used by Rodr\'\i{}guez \etal \cite{dani}: at a given streamwise station $x$, the radial variations of the global mode perturbation quantities are extracted. Since the eigenvector of the spatial local problem contains auxiliary variables $k\hat{u}_r$ and $k\hat{u}_x$ \citep[see][]{Lesshafft:2007p54}, these must also be added to the extracted slice of the global mode. This is accomplished by computing the streamwise derivative of the global $\hat{u}_r$ and $\hat{u}_x$ fields, and by augmenting the extracted vector $[\hat{u}_r,\hat{u}_x,\hat{p}]^T$ with $[-\i\p_x \hat{u}_r,-\i\p_x \hat{u}_x]^T$. This augmented global slice is finally projected onto the local modes, via scalar multiplication with the associated adjoint modes. As usual, the adjoint modes are normalised beforehand in such a way that their scalar product with the associated direct mode is unity, whereas the direct modes are by themselves scaled to have unit norm.

\begin{figure}
\centering
\includegraphics[width=\textwidth]{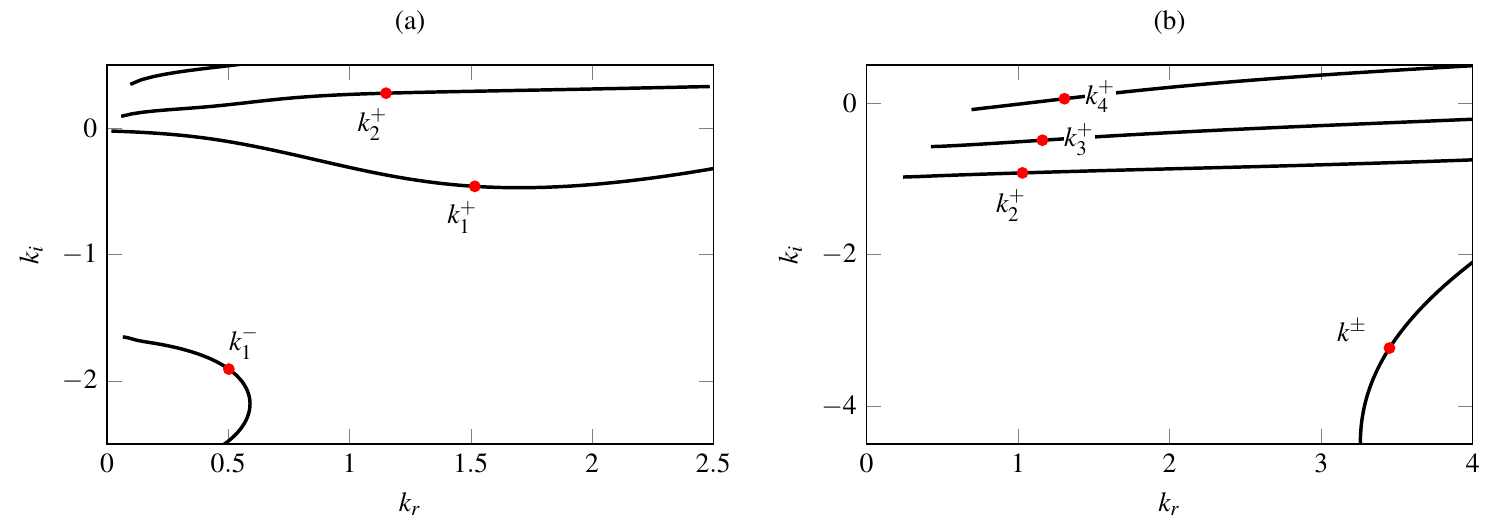}
\caption{Spatial eigenvalues from a local analysis of the parallel jet. Lines represent branches obtained by varying the real part of the frequency. \textit{a}) global mode 1: $\omega=1.111 - 0.052\i$; \textit{b}) global mode 2: $\omega=0.912 - 1.211\i$.}
\label{fig:localspectra}
\end{figure}

The spatial local spectrum of the parallel jet profile, shown in Fig.~\ref{fig:localspectra} for the frequency values of global modes 1 and 2, is composed of discrete $k^+$ and $k^-$ modes \cite{HM90}, which represent downstream- and upstream-propagating hydrodynamic perturbations inside and near the jet. The labels in Fig.~\ref{fig:localspectra} rank all $k^+$ and $k^-$ modes according to their spatial growth rate $-k_i$. 

\begin{figure}
\centering
\includegraphics[width=0.8\textwidth]{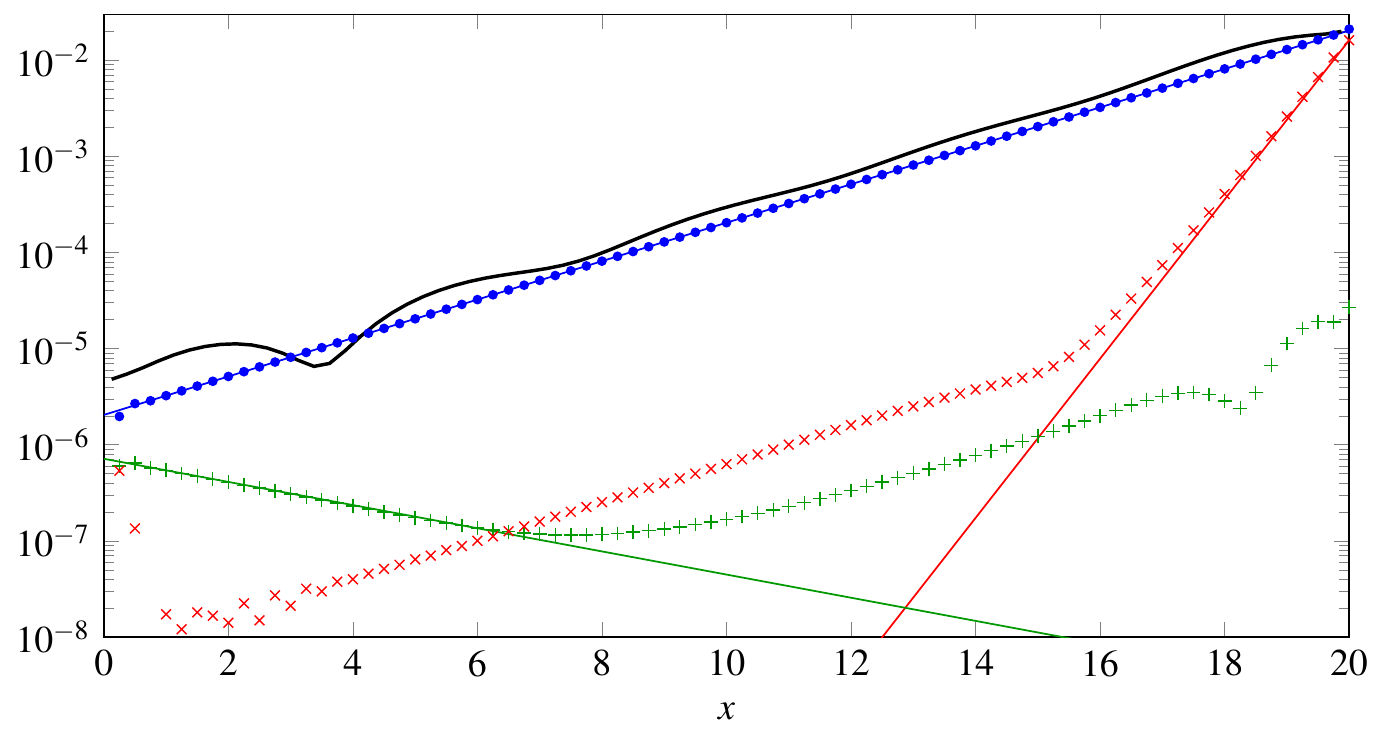}
\caption{Projection of global mode 1 onto spatial local modes of frequency $\omega=1.111 - 0.052\i$: absolute values of the projection coefficients as functions of $x$.
Legend: ({\color{blue}$\bullet$}) $k^+_1$;  ({\color{green!60!black}$+$}) $k^+_2$; ({\color{red}$\times$}) $k^-_1$; (\protect\rule[1.5pt]{4mm}{1pt}) global mode component $\hat{u}_x(0,x)$ on axis; colored lines indicate the growth rates according to local analysis. }
\label{fig:projection1}
\end{figure}

Figure \ref{fig:projection1} displays absolute values of the projection coefficients, obtained for global mode 1, pertaining to the three dominant local modes. These are the first two $k^+$ modes and the first $k^-$ mode. Blue symbols represent the $k^+_1$ mode (the only one displaying unstable spatial growth). Except very near the Dirichlet inlet, the streamwise variation of this local mode amplitude is perfectly exponential, with a spatial growth rate $0.4604$, as measured by a regression fit over $1\le x \le 19$. This value matches within $0.01\%$ the imaginary part of the local eigenvalue. The amplitude of the global mode component $\hat{u}_x$ on the jet centerline is shown as a black line for reference. 

\begin{figure}
\centering
\includegraphics[width=0.8\textwidth]{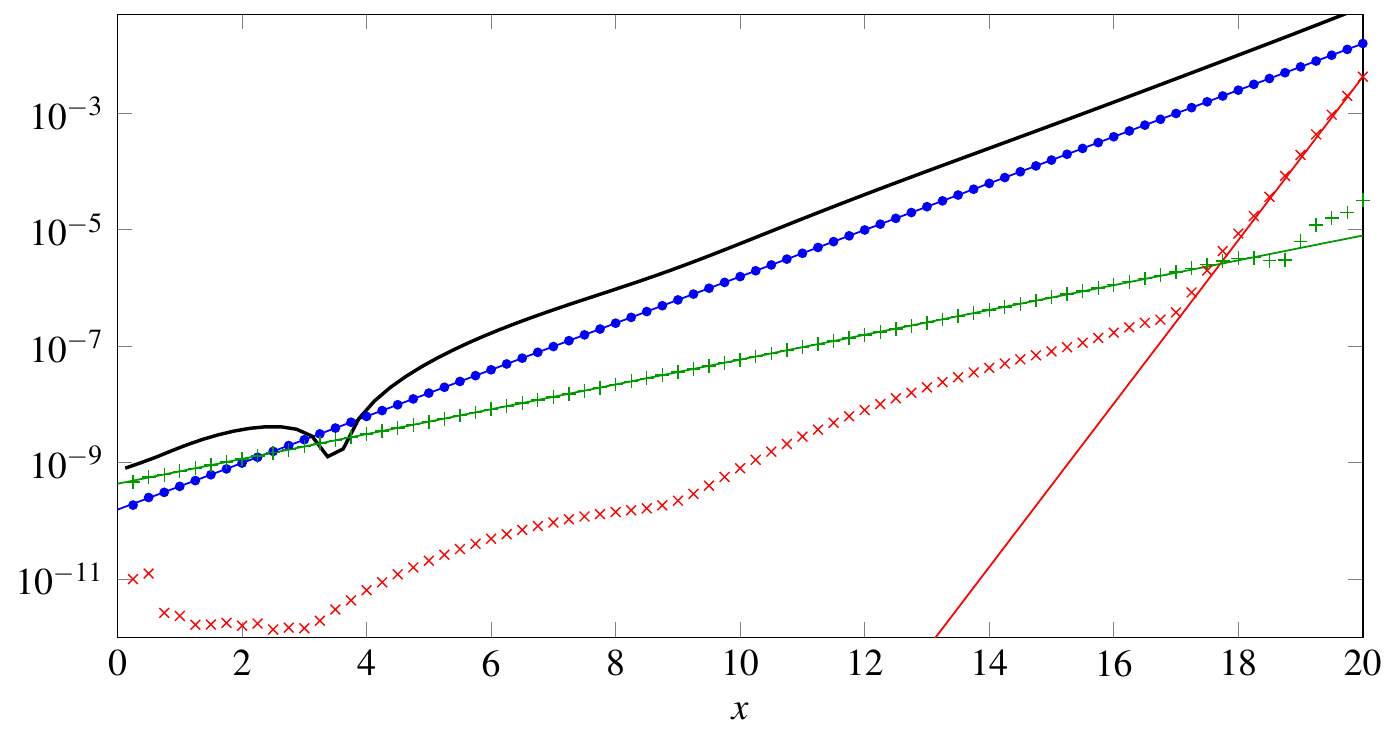}
\caption{Projection of global mode 2 onto spatial local modes of frequency $\omega=0.912 - 1.211\i$: absolute values of the projection coefficients as functions of $x$.
Legend: ({\color{blue}$\bullet$}) $k^+_2$;  ({\color{green!60!black}$+$}) $k^+_3$; ({\color{red}$\times$}) $k^\pm$; (\protect\rule[1.5pt]{4mm}{1pt}) global mode component $\hat{u}_x(0,x)$ on axis; colored lines indicate the growth rates according to local analysis. }
\label{fig:projection2}
\end{figure}

Green symbols in Fig.~\ref{fig:projection1} denote the amplitude of $k^+_2$, and red symbols the amplitude of $k^-_1$. Straight lines indicate the corresponding growth rate of the local eigenvalue for comparison. It is seen that both projections follow the amplitude variations expected from local analysis in a region close to one boundary, where their amplitude is maximal. Clearly, the $k^+_2$ mode originates at the upstream boundary, whereas the $k^-_1$ mode is forced at the downstream end. Farther away from those boundaries, both projections approximately follow the slope of the dominant $k^+_1$ branch. This behaviour results from imperfections in the numerical projection, which apparently only allows a clean distinction between these local modes down to amplitude ratios around $10^{-3}$ in the present setup. Higher spatial resolution in the local and global computations does not improve this threshold. Note that these three modes are highly non-orthogonal, which makes their distinction numerically delicate.

Overall, local mode contributions other than from the $k^+_1$ mode to the global mode 1 are tractable but rather negligible. Modulations of the centerline velocity perturbation (black line in Fig.~\ref{fig:projection1}) are instead attributable to global pressure modes, as will be shown later on.

Very similar results are obtained for the global mode 2, for which the local mode amplitudes are displayed in Fig.~\ref{fig:projection2}. However, the local spectrum in this case differs from that of mode 1, as the global frequency $\omega_2=0.912 - 1.211\i$ has an imaginary part below that of the absolute frequency $\omega_0=1.074-0.286\i$. Therefore, in the analysis of global mode 2, the local spatial modes are selected from a spectrum where pinching of the $k^+_1$ and the $k^-_1$ has already occurred (see Fig.~\ref{fig:localspectra}\textit{b}). In this setting, it is now the $k^+_2$ mode that displays the strongest downstream growth, followed by the $k^+_3$ mode. A mode from the mixed branch, formed from the $k^+_1$ and $k^-_1$ branches after pinching, is denoted $k^\pm$.

The global mode 2, as represented by a black line in Fig.~\ref{fig:projection2}, is clearly dominated by the $k^+_2$ mode (blue bullet symbols), but the $k^+_3$ and $k^\pm$ modes are again discernible down to amplitudes three orders of magnitude below $k^+_2$. It is not clear a priori how the $k^\pm$ mode is to be interpreted, in particular with regard to its up- or downstream propagation. However, the projection results plainly show that this mode is generated at the downstream end, from where it propagates upstream; thereby, it behaves as a mode of $k^-$ type. 

It is stressed that the analysis of spatial branches below the absolute growth rate is indeed meaningful in the present context. Spatial analysis below the absolute growth rate, i.e.~after pinching has taken place, is usually said to be in violation of temporal causality, formally expressed by the fact that no integration path can be found in the complex $k$-plane that separates $k^+$ and $k^-$ branches \citep{Huerre2000}. This argument however arises in the context of the asymptotic flow behaviour at long times, when indeed the system dynamics are determined by the absolute mode. In the same sense, a \emph{global} system is asymptotically determined by only the most unstable eigenmode (here: global mode 1). Notwithstanding, global eigenmodes with lesser growth rate do exist, and they are observable in the transient system dynamics. The spatial local modes used in the analysis of global mode 2 are justified in the same way.

In summary, the projections onto local spatial modes demonstrate that global mode 1 is supported by a local $k^+_1$ mode, whereas global mode 2 relies on a $k^+_2$ mode. In all likelihood, the same holds true for any global mode of the arc branch ($k^+_1$) and of the lower branch ($k^+_2$). Moreover, although not shown in Fig.~\ref{fig:globalspectrum}, further lower branches exist at lower growth rates $\omega_i<-1.5$, for which a match with higher local $k^+$ branches is anticipated. The link between the arc branch and the dominant $k^+$ mode has been pointed out in earlier studies \cite[for instance][]{Ehrenstein:2005p1026,Akervik:2008p1176,Garnaud:2013p1182}. However, the argument so far has one loose end: the presence of a local $k^+$ mode is contingent on it being forced upstream. The present results show that this forcing takes place immediately at the upstream boundary.
The essential ingredient that can give rise to a global mode with a $k^+$ wave is \emph{feedback} from downstream. 

%%%%%%%%%%%%%%%%%%%%%%%%%%%%%%%%%%%%%%%%%%%%%%%%

\subsection{Global pressure feedback}

The ellipticity of the global linear jet problem is contained in the pressure gradient and in the viscous terms. The latter will only be noticeable over distances much shorter than the numerical box length, and they are not considered in the following analysis. The pressure, as noted by Ehrenstein \& Gallaire \cite{Ehrenstein:2005p1026}, obeys a Poisson equation, which in the present case of parallel flow takes the form
\begin{equation}
\label{eqn:Poisson}
\Delta \hat{p} = -2\p_r U \p_x \hat{u}_r,
\end{equation}
with homogenous Dirichlet and Neumann conditions at the upstream and lateral boundaries, respectively, and with $\hat{p}=Re^{-1}\p_x \hat{u}_x$ at the outflow.

\begin{figure}
\centering
\includegraphics[width=\textwidth]{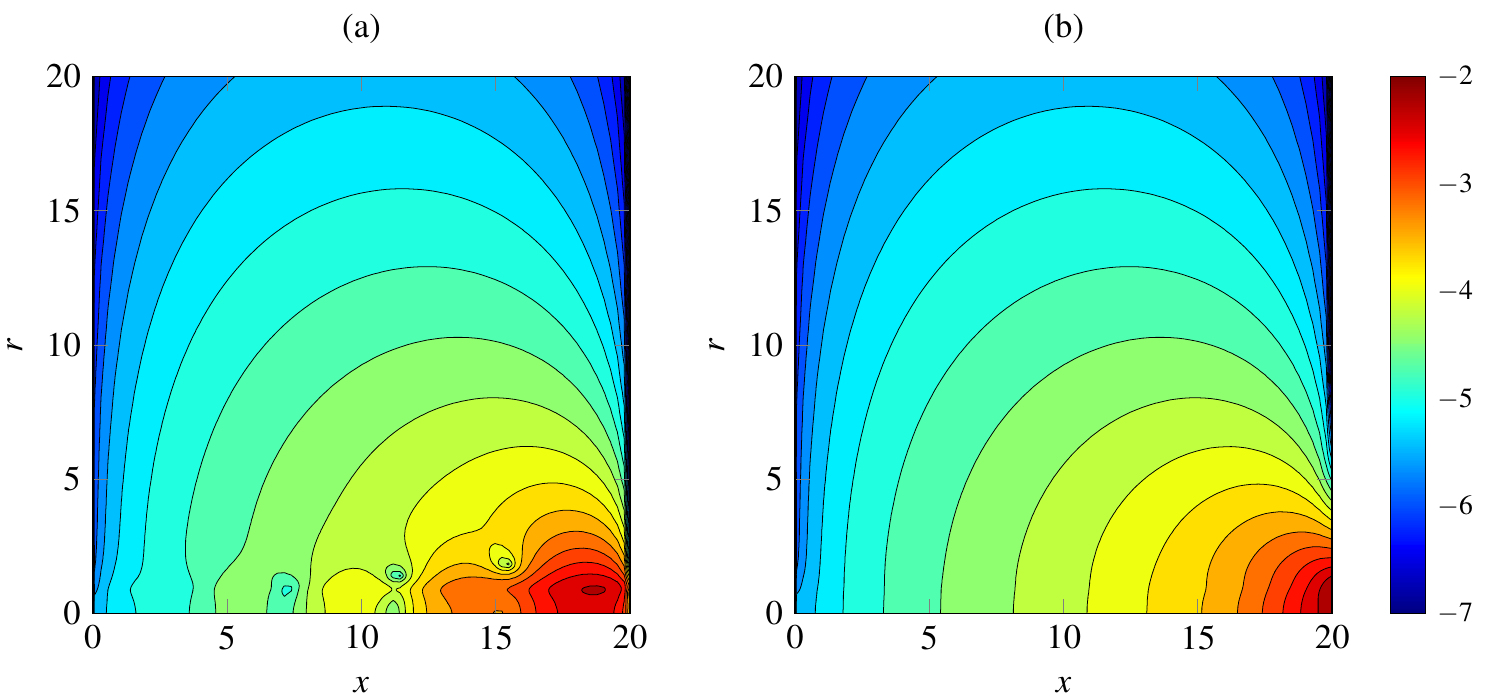}
\caption{Pressure perturbation fields of global mode 1. a) total pressure perturbation $\log_{10}|\hat{p}|$; b) pressure perturbation $\log_{10}|\tilde{p}|$, obtained by subtraction of the $k^+_1$ component.}
\label{fig:logpressure}
\end{figure}

The following analysis is restricted to the arc branch mode labelled `1' in Fig.~\ref{fig:globalspectrum}, but results for mode 2 are not fundamentally different. The pressure amplitude of mode 1 is shown in Fig.~\ref{fig:logpressure}(\textit{a}) as $\log_{10}|\hat{p}|$. Its structure is somewhat irregular in the downstream near-field region of the jet, yet the characteristic wavelength $2\pi/k_r=4.1$ of the $k^+_1$ mode is apparent.  The stress-free outflow boundary condition is seen to result in $\hat{p}\approx0$, and the pressure at $r\gtrsim 5$ is essentially a superposition of fundamental solutions of the homogeneous equation $\Delta \hat{p} = 0$, with Dirichlet conditions at the inflow and outflow. These are given by
\begin{equation}
\hat{p}_j = \left[A_j I_0\left(\frac{j\pi}{L}r\right) + B_j K_0\left(\frac{j\pi}{L}r\right)\right] \sin\left(\frac{j\pi}{L}x\right), \quad j \in \mathbb{N},
\end{equation}
where $I_0$ and $K_0$ are the modified Bessel functions of the first and second kind, respectively, and $L=20$ is the streamwise length of the numerical box. 
The $K_0$ functions are exponentially decaying in $r$, manifestly dominant in the present problem, whereas the $I_0$ functions grow exponentially in $r$. These only enter the global mode at very low amplitude ($A_j/B_j \ll 1$) in order to satisfy the Neumann condition at $r_{max}=50$. Only the $\hat{p}_1$ component is clearly visible in Fig.~\ref{fig:logpressure}\textit{a}, because it experiences the slowest radial decay, but a projection confirms that at least the first five $\hat{p}_j$ components enter the pressure field with comparable global norms.

It is known from the analysis in Sect.~\ref{sec:local} that the global mode involves a strong $k^+_1$ wave. In the present section, the complementary part is sought that may provide the upstream-reaching part of a feedback loop. Therefore, the $k^+_1$ component of the global mode is subtracted from the pressure field, using the already known projection coefficients shown in Fig.~\ref{fig:projection1}. This `stripped' pressure field $\tilde{p}$ is presented in Fig.~\ref{fig:logpressure}(\textit{b}), as $\log_{10}|\tilde{p}|$. A remarkably clean structure is recovered, suggestive of a solution of the Laplace equation $\Delta \tilde{p} = 0$ which is forced at the outflow boundary near the jet axis. A small inhomogeneity is also observed at the inflow near the the jet axis.

Note that the local $k^+_1$ mode by construction represents a particular solution of (\ref{eqn:Poisson}) in a domain of infinite streamwise extent. Within the limits of the simplifying assumption that the propagating perturbations in the region of $\p_r U\neq 0$ are indeed given by the $k^+_1$ mode alone, in the interior of the bounded domain, the stripped pressure field $\tilde{p}$ only needs to satisfy the inhomogeneities that arise from the boundary conditions. These inhomogeneities on both ends of the domain are thus coupled through the Laplace equation in $\tilde{p}$.

\begin{figure}
\centering
\includegraphics[width=0.7\textwidth]{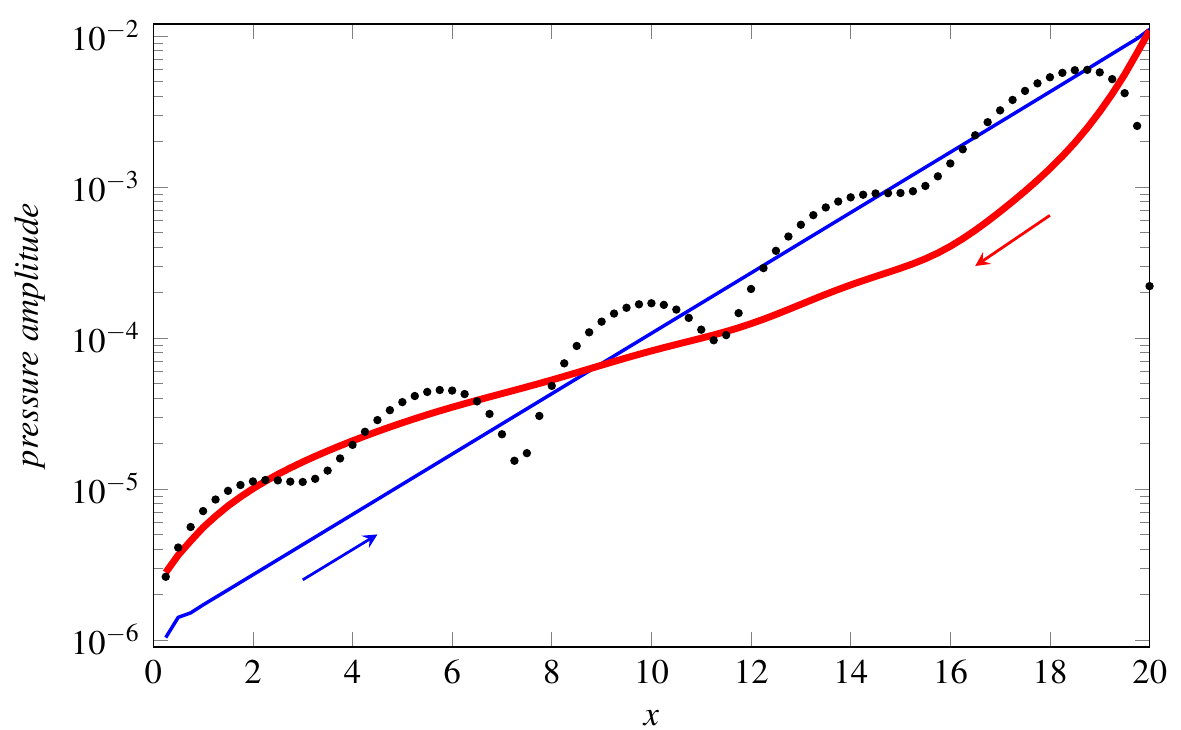}
\caption{Pressure signal amplitudes along $x$ at the critical point, $r_c=0.92$. ($\bullet$) total pressure $|\hat{p}|$; ({\color{blue}\protect\rule[1.5pt]{4mm}{0.75pt}}) pressure associated with $k^+_1$ wave; ({\color{red}\protect\rule[1.5pt]{4mm}{1.5pt}}) `stripped' pressure $|\tilde{p}|$. Note that all signals are numerically zero at $x=0$.}
\label{fig:feedbackloop}
\end{figure}

As a final plausibility check, the amplitudes of the supposed downstream-propagating and upstream-reaching components of the feedback loop are compared in Fig.~\ref{fig:feedbackloop} along a path at constant $r$. The radial position of the critical point of the local $k^+_1$ mode is chosen, $r_c=0.92$. A thin blue line represents the pressure amplitude of the $k^+_1$ mode component, according to the projection carried out in Sect.~\ref{sec:local}, a thick red line represents the amplitude of the pressure feedback signal $\tilde{p}$, and black symbols mark the amplitude of the total pressure field $\hat{p}$. The up- and downstream branches have approximately equal amplitude at the downstream boundary, nearly cancelling each other. Near the upstream boundary, the feedback signal is larger by about a factor 3 than the $k^+_1$ wave --- this appears reasonable in view of the assumption that the latter is forced by the former.

%%%%%%%%%%%%%%%%%%%%%%%%%%%%%%%%%%%%%%%%

\subsection{The influence of box length}
\label{sec:box}

\begin{figure}
	\centering
	\includegraphics[width=\textwidth]{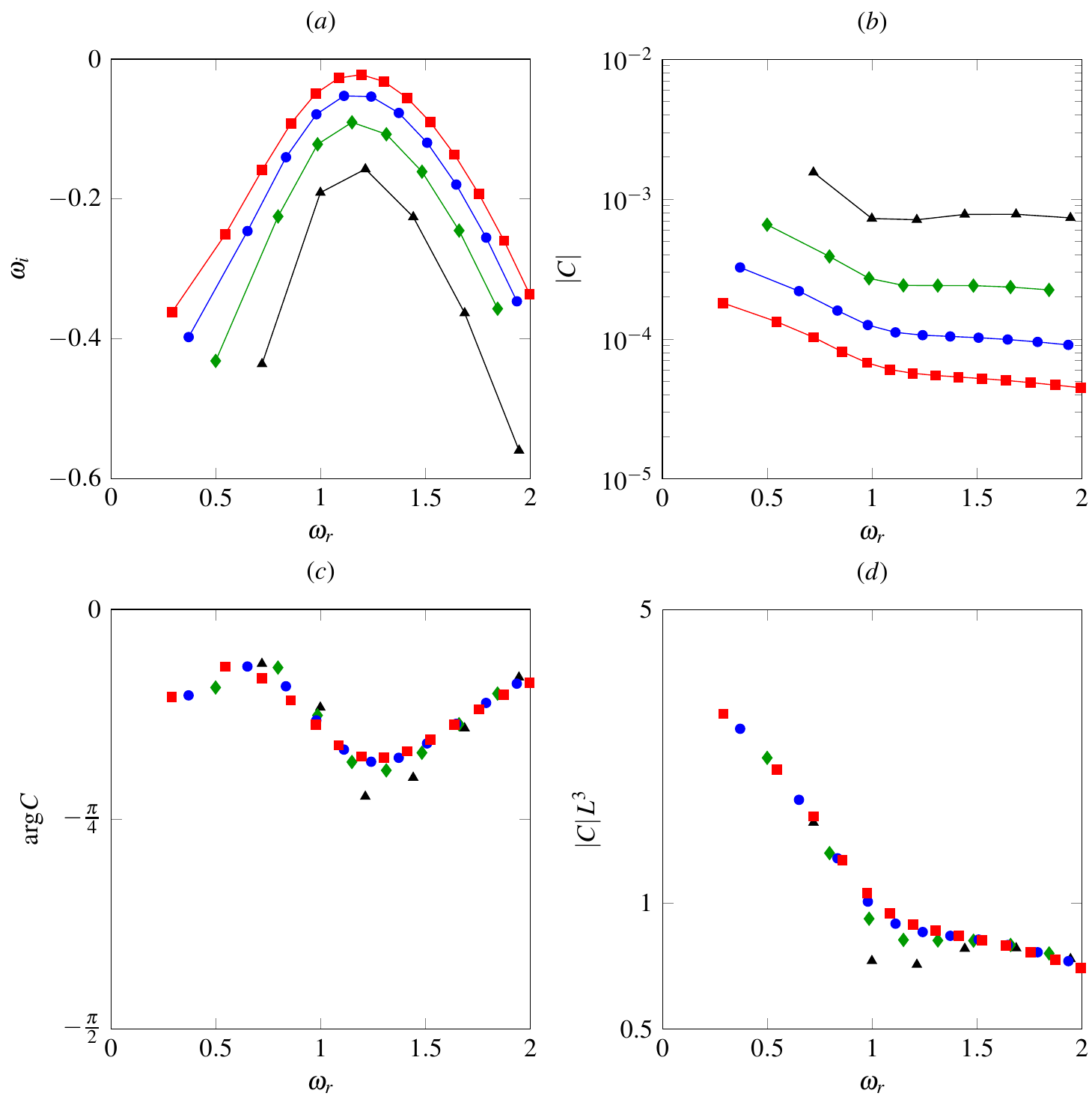}
	\caption{Influence of the numerical box length on eigenvalues. In all figures, different symbols denote different box lengths:  (\protect\raisebox{0.5pt}{\tiny\color{red}$\blacksquare$}) $L=25$; ({\color{blue}$\bullet$}) $L=20$; ({\scriptsize\color{green!60!black}$\blacklozenge$}) $L=15$; (\protect\raisebox{0.5pt}{\scriptsize$\blacktriangle$}) $L=10$. \textit{a}) Arc branch eigenvalues.  \textit{b}) Effective reflection coefficients according to (\ref{eqn:reflcoeff}), absolute values. \textit{c}) Effective reflection coefficients, phase. \textit{d}) Effective reflection coefficients, absolute value scaled with $L^3$.
	}
	\label{fig:boxsize}
\end{figure}

Arc branch modes in the literature are consistently found to be sensitive to the length of the numerical box. In some instances \citep[e.g.][]{Garnaud:2013p1182}, a longer box yields eigenvalues with lower growth rates, in other instances \citep[e.g.][]{Ehrenstein:2005p1026} the opposite effect is observed. If the present feedback model is correct, it should allow an estimation of the influence of the numerical box length on the arc branch growth rates.

Eigenmodes of the parallel jet (\ref{eqn:baseflow}) have been computed in numerical domains of streamwise lengths $L=10,15$ and 25, for comparison with the standard configuration $L=20$; in all cases, the radial discretisation and the constant step size $\Delta x$ are unchanged. Resulting modes of the arc branch are displayed in Fig.~\ref{fig:boxsize}(\textit{a}). As $L$ is increased, the entire branch is seen to shift to higher growth rates, and the modes are more densely spaced.

The analysis so far suggests that a downstream-convecting $k^+$ mode and the elliptic pressure field are coupled in small regions near the inflow and outflow boundaries. In analogy with the Ginzburg--Landau model discussed in Sect.~\ref{sec:GL}, an effective reflection coefficient $C$ may be defined, which relates the forcing of the $k^+$ mode at the inflow to the $k^+$ amplitude at the outflow, thus lumping the narrow interaction regions into singular actuator and sensor positions in $x$. The effective forcing at the Dirichlet inflow boundary is modelled as
\begin{equation}
\label{eqn:reflcoeff}
-\i\omega\hat{q} = C\hat{q}\,\text{e}^{\i kL} \quad \rightarrow \quad C = -\i\omega \text{e}^{-\i kL},
\end{equation}
where $\omega$ is the global eigenmode frequency, and ($k,\hat{q}$) denote the relevant local $k^+$ mode.  The coefficient $C$ includes the effects of coupling on both ends of the domain, as well as the upstream decay of the pressure signal, seen in Fig.~\ref{fig:feedbackloop}.

Numerically obtained values of $C$ are reported in figures \ref{fig:boxsize}(\emph{b},\emph{c}), in terms of their absolute value and their phase, for the arc branch in all four numerical domains. In each domain, the reflection coefficient decreases with real frequency, but only weakly so for frequencies larger than one. This effect may be caused by the slower radial decay of $\hat{q}$ at low frequencies, resulting in radially more extended interaction regions on both domain ends. The phase of $C$ varies only slightly with frequency, and it is independent of box length. 

If the reflection coefficients are scaled with the cube of the box length, as shown in Fig.~\ref{fig:boxsize}(\emph{d}), they neatly collapse onto one curve. Consequently, the effective feedback imparted by boundary reflections decreases with the domain length as $L^3$. This scaling factor indicates that the relevant component of the pressure signal, which couples the downstream with the upstream end of the $k^+$ wave, is of a \emph{quadrupole} type (signals from monopole and dipole sources at $x=L$ would decay as $L$ and $L^2$, respectively).
%If the outflow condition near the axis is modelled as a point source of $\Delta\tilde{p}=0$ in an \emph{unbounded} domain, the resulting pressure field can be expressed in terms of multipole components,
%\begin{align}
%\text{monopole:~~} & \tilde{p}_m(\xi,\theta) \sim \xi^{-1},\\
%\text{dipole:~~} & \tilde{p}_d(\xi,\theta) \sim \xi^{-2} \cos\theta,\\
%\text{quadrupole:~~} & \tilde{p}_q(\xi,\theta) \sim \xi^{-3}(\cos^3\theta-1),
%\end{align}
%where the spherical coordinates $\xi$ and $\theta$ denote the distance from the source and the inclination angle with respect to the jet axis.

It is then straightforward to interpret the effect of an increased numerical box length on a given arc branch mode: if the exponential growth of the dominant $k^+$ mode over the added streamwise interval is larger than the algebraic decay of the upstream-reaching pressure signal, over the same added distance, then the spurious inlet forcing will increase in strength, resulting in a higher growth rate $\omega_i$. If the $k^+$ mode in the added region is locally stable, then the inlet forcing will decrease, and the global growth rate will be lower as a result. 

This interpretation appears to be consistent with the two cited examples: the boundary layer investigated by Ehrenstein \& Gallaire \cite{Ehrenstein:2005p1026} is locally unstable at the outflow, as confirmed by the authors, and longer domain sizes are found to result in higher global growth rates. In contrast, the rapidly spreading jet considered by Garnaud \etal \cite{Garnaud:2013p1182} is locally stable at the outflow with respect to axisymmetric perturbations, and indeed lower global growth rates are obtained for the arc branch in longer domains. This criterion may of course be frequency-dependent: certain (complex) frequencies may be locally stable at the outflow while others are unstable, and opposite trends ought to be observed in these frequency ranges. This seems indeed to be the case in the $Re=360$ setting of Coenen \etal \cite{Coenen2017}.
%\cite{Lesshafft2014}.

%It is not the purpose of this study to identify numerical boundary conditions for minimised feedback. However, the above discussion suggests that \emph{absorbing layers} \citep{C04} near the outflow should in all circumstances be effective, provided that strong enough damping is applied in order to achieve significant spatial decay of perturbations. Such a treatment may be chosen in cases where it is desirable to lower the temporal growth rates of arc branch modes, especially if these turn out to be unstable \citep[see the discussion by][]{Kurz2016}. However, it will in general not be possible to make these modes disappear, or to bring them to `convergence' in a physically meaningful way.

%%%%%%%%%%%%%%%%%%%%%%%%%%%%%%%%%%%%%%%%%%%%

\section{Possible remedies, tested for a non-parallel jet}
\label{sec:nonpar}

While the preceding analyses of the Ginzburg--Landau equation and of the parallel jet served to \emph{characterise} the global feedback mechanism due to domain truncation, the question how such feedback may be \emph{reduced} is addressed in this section for the example of a spatially developing jet. With this choice, the results from the previous section can be largely transferred to the new setting.

Each type of open flow may present particular problems in view of domain truncation. In favourable configurations, the numerical boundaries can be placed in stable flow regions, as for instance in a uniform flow upstream of a solid obstacle, like a cylinder. In other cases, the need for truncation may be avoided by prescribing consistent physical boundaries, like confining solid walls, which are then part of the flow configuration.

Jets belong to a more problematic category. They are created by some upstream source of momentum, and it is in general not feasible to include the entire upstream apparatus (fan, chamber, nozzle, etc.) in the calculations, which would furthermore defeat the purpose of any generic description of jet dynamics. The formulation of upstream flow and boundary conditions is therefore necessarily imperfect with respect to any flow realisation, as potentially important regions are not accounted for.

\begin{figure}
	\centering
	\includegraphics[width=\textwidth]{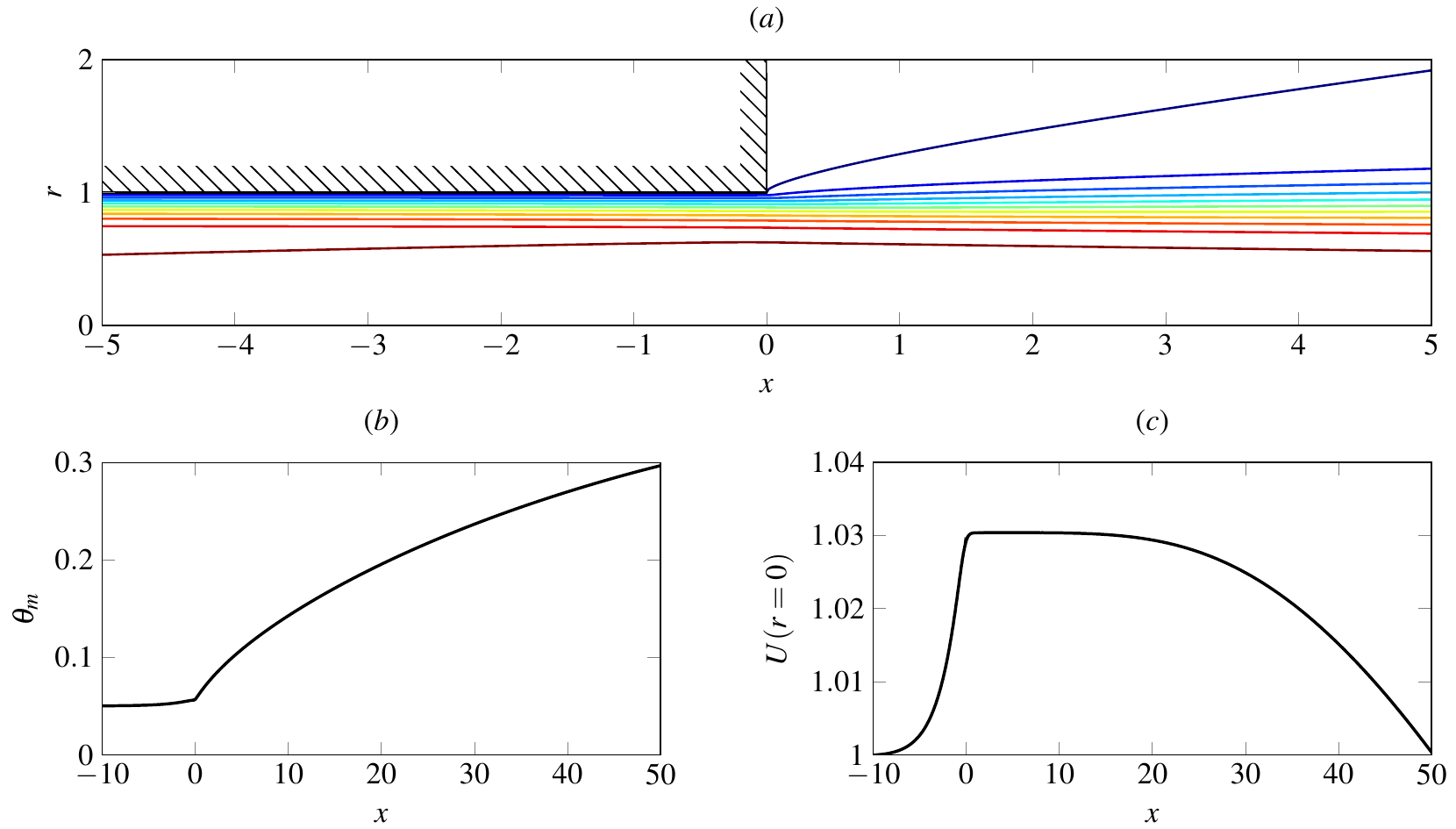}
	\caption{Steady state of a spatially developing jet, used as a base flow in all calculations of Sect.~\ref{sec:nonpar}. \textit{a}) Contours of axial velocity near the orifice at $x=0$. Contour values are distributed between 0 and 1, with constant spacing of 0.1. The base flow is computed on a larger domain, with $-10\le x\le 50$ and $0\le r \le 50$. \textit{b}) Streamwise development of the shear layer momentum thickness. \textit{c}) Streamwise development of the centreline velocity.}
	\label{fig:baseflow}
\end{figure}

For the present study, the jet is modelled as issuing from an orifice in a solid wall, very similar to the configuration of Garnaud \textit{et al.}\cite{Garnaud:2013p1182}. Upstream of the orifice, the flow develops in a straight circular pipe, from where it exits with a fairly thin boundary layer. The base flow is computed as a steady, axisymmetric solution of the incompressible Navier--Stokes equations (\ref{eqn:nonlineqns}), with an inflow condition
\begin{equation}
	U(r)=\tanh \,\left[\frac{5}{2}\left(\frac{1}{r}-r\right)\right]
\end{equation}
imposed at $x=-10$. Newton--Raphson iterations are performed in order to converge to a steady flow state. The numerical domain for these base flow calculations is truncated downstream at $x_{max}=50$ and radially at $r_{max}=50$, where stress-free conditions are applied. The Reynolds number is set to 1000 in the free jet. However, in order to maintain a thin shear layer at the orifice, this value is varied exponentially inside the pipe, between $Re=10^5$ at $x=-10$ and $Re=10^3$ at $x\ge -0.2$, in the base flow calculation. The resulting flow field near the orifice at $x=0$ is represented in Fig.~\ref{fig:baseflow}, together with the characteristic streamwise variations of the shear layer momentum thickness $\theta_m$ and of the centreline velocity. %The maximum absolute growth rate occurs near $x=0.01$, where $\omega_0=1.610-0.112i$, therefore the flow is convectively unstable everywhere.

Eigenmode calculations are carried out on smaller domains, where portions of the base flow are cropped at the inflow, the outflow, or both, in order to probe the effect of domain truncation. All configurations are listed in table \ref{tab:configs}. The Reynolds number in these calculations is set to 1000 throughout the domain.

\begin{table}
	\centering
	\begin{tabular}{l|rrll}
		~ & $x_{min}$ & $x_{max}$ & upstream BC & absorbing layer \\
		\hline
		case 1 & 0 & 40 & Dirichlet & no \phantom{\Large A} \\
		case 2 & $-5$ & 40 & Dirichlet & no \\
		case 3 & $-10$ & 40 & Dirichlet & no \\
		case 4 & 0 & 40 & stress-free & no \\
		case 5 & $-5$ & 40 & stress-free & no \\
		case 6 & $-10$ & 40 & stress-free & no \\
		case 7 & $-5$ & 40 & Dirichlet & yes, $\lambda_{max}=0.5$ \\
		case 8 & $-5$ & 50 & Dirichlet & yes, $\lambda_{max}=0.75$ \\
	\end{tabular}
	\caption{Numerical configurations of the various eigenmode calculations discussed in this section.} 
	\label{tab:configs}
\end{table}

\begin{figure}
	\centering
	\includegraphics[width=0.9\textwidth]{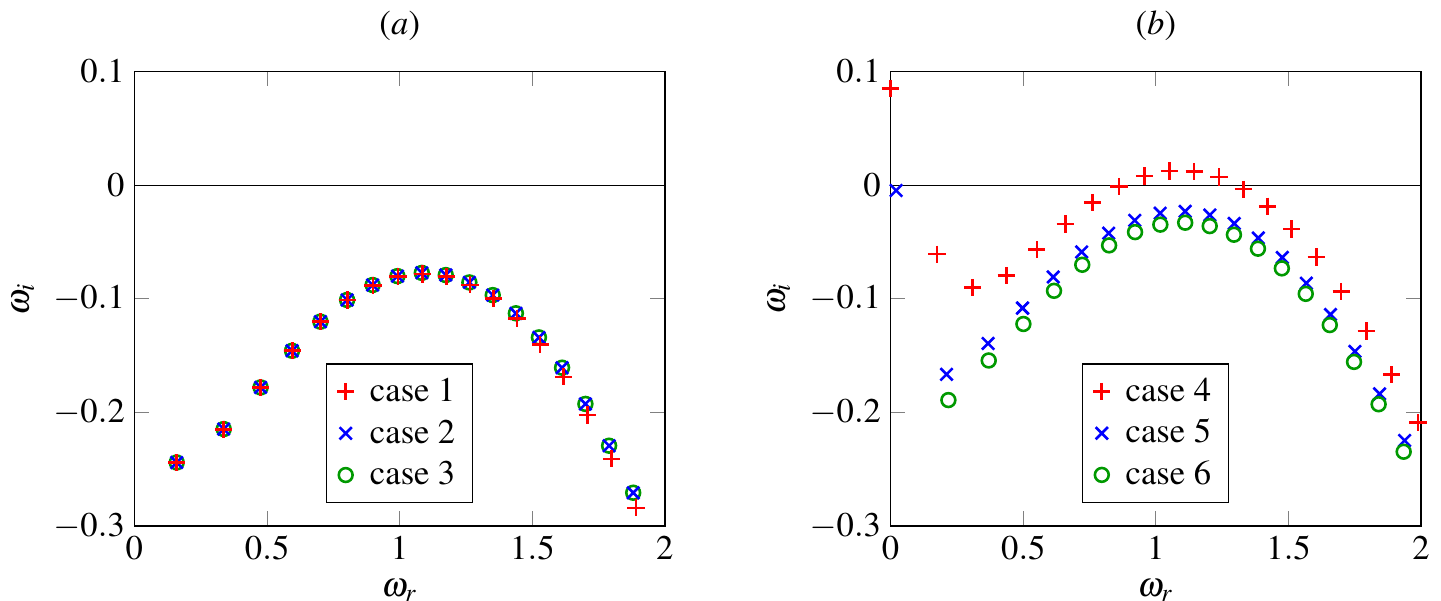}
	\caption{Arc branch spectra obtained in calculations with (\textit{a}) Dirichlet and (\textit{b}) stress-free upstream boundary conditions.}
	\label{fig:evals_upstreambc}
\end{figure}

The influence of \emph{upstream} truncation on eigenmodes is considered first. Figure \ref{fig:evals_upstreambc}a shows spectra for three different domains, with homogeneous Dirichlet conditions (\ref{eqn:bcs}a) imposed at $x_{min}=(0,-5,-10)$, respectively. The downstream end of the numerical domain is placed at $x_{max}=40$, where stress-free conditions (\ref{eqn:bcs}b) are applied in all three cases. Unconverged and lower-branch eigenvalues are not shown here and in the following for clarity; the criterion for convergence is that an eigenvalue could be reproduced within three-digit accuracy using two different shift values. 

It is seen from Fig.~\ref{fig:evals_upstreambc}a that the arc branch
is quite insensitive to the position of the upstream Dirichlet boundary. Even a full truncation of the upstream pipe (case 1) only results in a slight stabilisation at high frequencies. Eigenvalues obtained with pipe lengths of 5 and 10 radii (cases 2 and 3) are virtually identical. Stress-free inflow conditions (cases 4, 5 and 6, spectra shown in Fig.~\ref{fig:evals_upstreambc}b) are found to perform less favourably. When applied at $x=0$, these conditions give rise to global instability; the inclusion of portions of the pipe has a stabilising effect, but even with $x_{min}=-10$ the growth rates along the entire arc branch are still significantly higher than those obtained with Dirichlet conditions.

It is to be expected that global pressure fluctuations induce a generation of vorticity waves at the solid corner at $x=0$, and the independence of the eigenvalues in cases 2 and 3 (Fig.~\ref{fig:evals_upstreambc}a) on the upstream boundary position suggests that this physical effect is dominant over spurious pressure-vorticity coupling at $x_{min}$. It remains to be determined to what extent the downstream boundary conditions emit spurious pressure signals, and how these may be reduced. 

\begin{figure}
	\centering
	\includegraphics[width=0.75\textwidth]{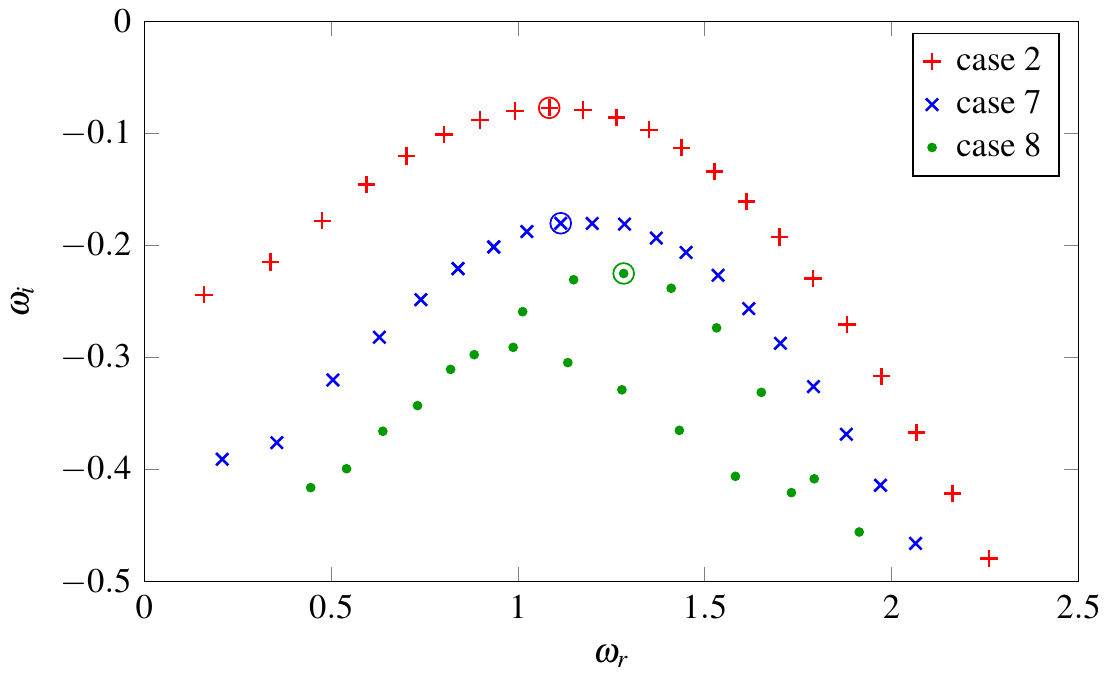}
	\caption{Arc branch spectra obtained in calculations with ({\color{blue}$\times$},{\color{green!60!black}$\bullet$}) and without ({\color{red}$+$}) absorbing layer. Circles mark the individual modes for which eigenfunctions are shown in Fig.~\ref{fig:evecs_downstreambc}.}
	\label{fig:evals_downstreambc}
\end{figure}
\begin{figure}
	\centering
	\includegraphics[width=0.75\textwidth]{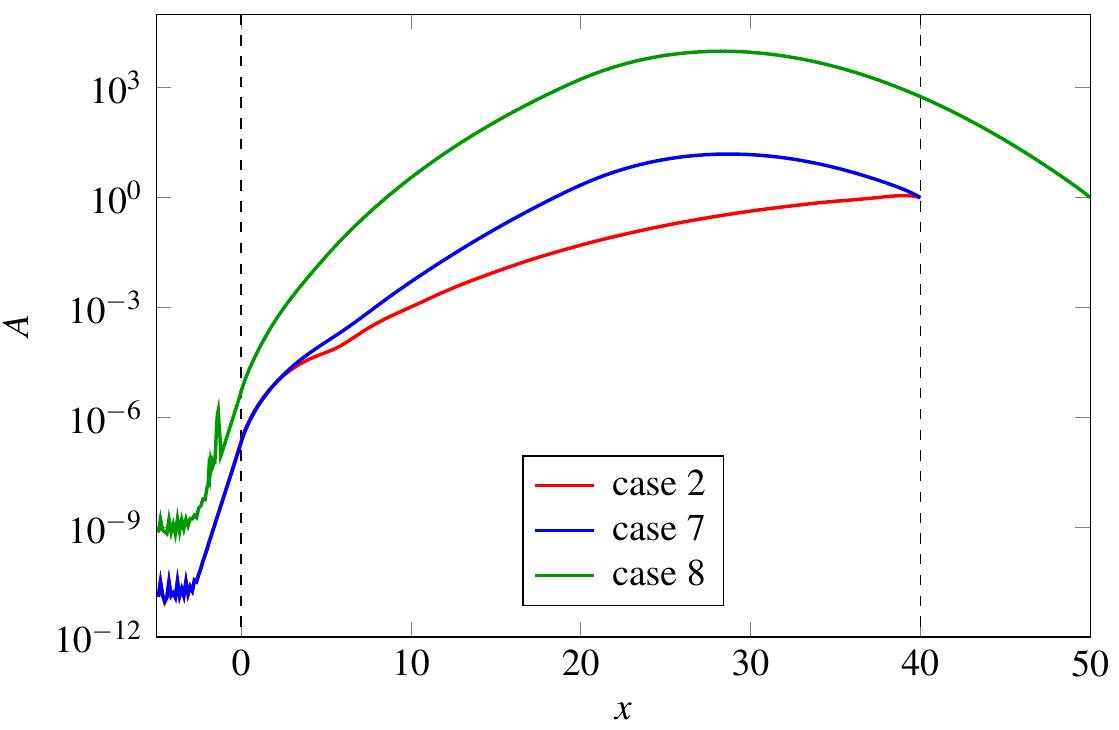}
	\caption{Amplitude variations in $x$ of selected eigenfunctions, as marked in Fig.~\ref{fig:evals_downstreambc}. The amplitude is defined in Eq.~(\ref{eqn:amplitude}).}
	\label{fig:evecs_downstreambc}
\end{figure}
Garnaud \textit{et al.~}\cite{Garnaud:2013p1182} tested stress-free and convective outflow conditions in an almost identical jet configuration, and found no significant difference in the resulting spectra. In the present study, the potential of \emph{absorbing layers} \citep{C04} for a stabilisation of the arc branch is examined. To this end, an artificial damping term is added to the linear perturbation equations, with the purpose of reducing perturbation amplitudes before they reach the numerical boundary at $x_{max}$. If the original global eigenvalue problem is written
\begin{equation}
-i\omega \mathbf{B} \hat{q} = \mathbf{L} \hat{q},
\end{equation}
then the eigenvalue problem with absorbing layer is defined as
\begin{equation}
-i\omega \mathbf{B} \hat{q} = \left[\mathbf{L} - \lambda(x)\mathbf{B}\right]\hat{q}.
\end{equation}
For cases 7 and 8 (see table \ref{tab:configs}), the damping parameter is prescribed as
\begin{equation}
\lambda(x) = \left\{
\begin{array}{rl}
0 & \text{~~~for~~} x \le 19,\\
0.00625x^2 - 0.2375x + 2.25625 & \text{~~~for~~} 19<x<21,\\
0.025(x-20) & \text{~~~for~~} x \ge 21.
\end{array}\right.
\end{equation}
This variation of $\lambda$ is continuous in its first derivative, and it has maximum values 0.5 and 0.75 at $x_{max}=40$ and $x_{max}=50$, respectively.

The effect of absorbing layers on the eigenvalue spectrum is documented in Fig.~\ref{fig:evals_downstreambc}. Damping in the downstream region $19\le x \le 40$ (case 7) is seen to reduce the growth rates of arc branch modes significantly (compared to case 2). Eigenfunctions of selected modes, marked by circles in Fig.~\ref{fig:evals_downstreambc}, are compared in Fig.~\ref{fig:evecs_downstreambc}. An integral amplitude measure is defined as
\begin{equation}
A(x)=\left[\int_0^1 \left( |\hat{u}_r|^2 + |\hat{u}_x|^2 \right) r\, \text{d}r \right]^\frac{1}{2},
\label{eqn:amplitude}
\end{equation}
and the amplitude curves in Fig.~\ref{fig:evecs_downstreambc} are normalised  with respect to their values at the outflow boundary for comparison. While the red curve (case 2, no absorbing layer) shows a monotonic growth of perturbation amplitude throughout the domain, similar to the parallel jet results in figures \ref{fig:projection1} and \ref{fig:projection2}, the blue curve (case 7) reaches a maximum inside the absorbing layer and subsequently decays in $x$. Remarkably, the ratio $A(x_{max})/A(0)$ is identical in both cases, which is consistent with the interpretation of these modes as being the result of spurious feedback from the outflow, in the same way as detailed in Sect.~\ref{sec:jet} for the parallel jet. The absorbing layer reduces the gain of the $k^+$ hydrodynamic branch of the feedback loop, and thereby the temporal modal growth rate.

Based on the discussion in Sect.~\ref{sec:jet}, stronger artificial damping of the hydrodynamic branch ought to lead to ever smaller modal growth rates; longer domains should furthermore lead to a reduced amplitude of the incident pressure signal upstream, due to its cubic decay. Both conditions are combined in case 8 (see table \ref{tab:configs}), of which results are included in figures \ref{fig:evals_downstreambc} and \ref{fig:evecs_downstreambc}. Indeed, the temporal growth rates are further reduced, and the spatial decay of the eigenfunction amplitude in the absorbing layer is more pronounced than in the case 7.

However, the shape of the arc branch in case 8 displays some differences with respect to all other cases, and similar mode patterns have been reported from jet calculations on long domains by Coenen \etal~\cite{Coenen2017}, where no artificial damping was applied. In the vicinity of the least stable mode (circled in Fig.~\ref{fig:evals_downstreambc}), a regular spacing in $\omega_r$ is still observed, but with larger distances between consecutive modes than in cases 2 and 7. The corresponding amplitude function in Fig.~\ref{fig:evecs_downstreambc} shows that the ratio $A(x_{max})/A(0)$ is smaller than in the two other cases, whereas the cubic decay of the reflection coefficient with box length, as described in Sect.~\ref{sec:jet}, should have resulted in a larger ratio. Furthermore, stronger damping through higher values of $\lambda$, tried in test calculations that are not shown here, does not decrease the maximum growth rate much further, but it quickly leads to ill-conditioned system matrices. Case 8 seems indeed to mark an efficiency limit of absorbing layers for the stabilisation of the arc branch.

The above observations, in particular the increased spacing of modes in $\omega_r$, suggest that pressure feedback in the strongly damped case 8 may originate from the interior of the domain, possibly from the location of the amplitude maximum. Such feedback may be spurious, in the sense of Heaton \etal~\cite{heaton}, or it may be physical. This hypothesis is suggested for further examination.

%%%%%%%%%%%%%%%%%%%%%%%%%%%%%%%%%%%%%%%%%%%%
\section{Conclusions}

A family of linear instability eigenmodes, named the arc branch, has been analysed in view of its physical or numerical origin. Branches of this type have previously been observed in a large variety of open flows. All results presented herein lead to the conclusion that arc branch modes in incompressible flow calculations are an artifact of domain truncation, due to spurious pressure feedback between numerical inflow and outflow boundaries.

A Ginzburg--Landau model was investigated first, because in this setting the effect of explicitly prescribed feedback between a downstream sensor and an upstream actuator could be examined without ambiguity. In the presence of feedback, a branch of eigenmodes was observed to arise that exhibits all typical characteristics of the arc branch described by Coenen \etal \cite{Coenen2017} for the example of light jets. These modes align with a contour of the pseudospectrum, spaced at regular intervals of the real frequency, and their spatial eigenfunction is characterised by an integer number of wavelengths between actuator and sensor locations. As the strength of the feedback is increased, the branch moves steadily upward to higher growth rates in the complex frequency plane.  Eigenmodes of the feedback-free system that lie below the arc branch are no longer detectable. An important observation is that this arc branch in the Ginzburg--Landau equation with feedback has no counterpart in the feedback-free spectrum: these eigenmodes are not merely \emph{affected} by the presence of feedback, indeed without it they \emph{do not exist}.

An incompressible parallel jet in a truncated domain was examined next. No explicit feedback was prescribed in this setting, but a similar arc branch of eigenmodes was nonetheless found to dominate the spectrum. A lower branch of eigenmodes with stronger temporal decay was also described, which may correspond to subdominant branches observed in wakes and boundary layers. Both the arc-branch and the lower-branch modes in the parallel jet were shown to be composed primarily of one downstream-propagating $k^+$ wave, as computed from a local spatial analysis. Each of the two branches involves a different $k^+$ mode. This observation raises the question how a $k^+$ wave is generated at the upstream boundary of the numerical domain. The analogy with the Ginzburg--Landau model from Sect.~\ref{sec:GL} suggests feedback from downstream.

The most plausible mechanism for global feedback in a truncated domain is the generation of pressure perturbations due to the artificial downstream boundary condition, as described by Buell \& Huerre\cite{BuellHuerre}. The pressure field of the least stable arc branch mode was decomposed into one portion associated with the prominent $k^+$ wave, which cannot be involved in upstream feedback, and into a residual part that is essentially governed by a Laplace equation forced at the domain boundaries. The latter appears to be strongly dominated by a source region at the outflow, near the jet axis. A numerical evaluation of effective reflection coefficients, in analogy with the Ginzburg--Landau model of  Sect.~\ref{sec:GL}, established as a principal result that the strength of upstream feedback decays with the numerical box length to the third power. This scaling is indicative of a spurious pressure quadrupole situated at the outflow. The algebraic nature of the pressure decay allowed a prediction of the effect of box length variations on arc-branch growth rates, depending on the local stability or instability of the flow, apparently consistent with common observations.
While this study has been limited to incompressible flow settings, similar mechanisms will also be present in compressible calculations, with the difference that spurious pressure signals are then  propagated by a wave equation. This may in fact result in stronger feedback, as the far-field acoustic pressure only decays with the first power of the distance from the source.

Finally, possible strategies for a reduction of spurious pressure feedback have been examined for the example of a spreading jet. On the one hand, upstream boundary conditions must be chosen that minimise the unphysical conversion of pressure feedback from downstream to vortical perturbations. It has been found, in this particular flow example, that Dirichlet conditions perform much better than stress-free conditions in this regard. On the other hand, the generation of spurious pressure signals from the outflow boundary must be reduced. It has been demonstrated that artificial damping in a downstream absorbing layer provides an efficient means to achieve this. An advantage of this technique is that it is straightforward to implement in any usual open shear flow problem. It has been noted, however, that increasingly strong damping does not reduce the growth rate of the arc branch to arbitrarily low levels.

\begin{acknowledgements}
Discussions with Xavier Garnaud and Wilfried Coenen greatly helped shape the ideas presented in this paper.
The study was supported by the Agence Nationale de la Recherche under the Cool Jazz project, grant number ANR-12-BS09-0024. 
\end{acknowledgements}

\bibliographystyle{plain}

\end{document}